\PassOptionsToPackage{table}{xcolor}
\documentclass[sigconf]{acmart}
\usepackage{popets}
\usepackage{algorithm}
\usepackage{algpseudocode}
\usepackage{booktabs, array, colortbl}
\usepackage{xcolor}
\usepackage{tikz, amsmath}
\usepackage[most]{tcolorbox}
\usepackage{enumitem}
\usepackage{afterpage}

\AtEndPreamble{%
  \theoremstyle{acmdefinition}
  \newtheorem{goal}{Goal}
}

\newcolumntype{C}[1]{>{\centering\arraybackslash}p{#1}}

\definecolor{enrollfill}{RGB}{204, 229, 255}
\definecolor{empfill}{RGB}{204, 255, 204}
\definecolor{outfill}{RGB}{255, 229, 204}
\usetikzlibrary{fit, backgrounds, arrows.meta, positioning, calc}
\tikzset{
  innerbox/.style={draw, rectangle, minimum width=1.8cm, minimum height=1.8cm},
  outerbox/.style={draw, rectangle, line width=1.4pt, inner sep=8pt},
  colhead/.style={draw, rectangle, minimum width=1.15cm, minimum height=0.6cm,
                  align=center, font=\small\ttfamily},
  fieldbox/.style={draw, rectangle, minimum width=1.15cm, minimum height=0.6cm,
                   align=center, font=\small\ttfamily},
  jtable/.style={draw, rectangle, minimum width=3.6cm, minimum height=1.5cm},
  ann/.style={font=\scriptsize\sffamily},
}

\setcopyright{popets}
\copyrightyear{YYYY}

\acmYear{YYYY}
\acmVolume{YYYY}
\acmNumber{X}
\acmDOI{XXXXXXX.XXXXXXX}
\acmISBN{}
\acmConference{Proceedings on Privacy Enhancing Technologies}
\DeclareMathOperator{\prov}{prov}

\begin{document}

\title{Proof-of-Retention: A Framework for Auditable Cross-Organization Data Sharing}

\author{Kyle MacMillan}
\affiliation{%
\institution{The University of Chicago}
\city{Chicago}
\country{USA}}
\email{macmillan@uchicago.edu}

\author{Sanjay Krishnan}
\affiliation{%
\institution{The University of Chicago}
\city{Chicago}
\country{USA}}
\email{skr@uchicago.edu}

\renewcommand{\shortauthors}{MacMillan et al.}

\begin{abstract}
The rapid adoption of AI across industries for (e.g.) fine-tuning and analytics
has accelerated the need for high-quality data. To satisfy this demand, public
and private entities will buy and sell data with other organizations. But such data
sharing can and does violate privacy norms and laws. EU and American lawmakers
have endeavored to control data sharing, restricting what data may be shared
with whom, and under what circumstances. Unfortunately, accurately assessing
compliance with new regulatory regimes remains a challenge, as \textit{data
provenance}, that is, metadata about data sharing, is rarely preserved. And even
when provenance information is retained, unilateral changes to one party's
database can render the provenance stale.

To fill this gap, we present \textit{Proof-of-Retention} a novel framework and
interactive protocol that enforces auditable data sharing. Our framework
requires each party involved in data sharing to retain a subset of information
associated with each data exchange, and provides a mechanism for auditors to verify that the
parties have indeed retained that information. Our approach combines techniques
from database systems, including query witness generation, with cryptographic
proof-of-possession protocols to provide retention guarantees. The framework
enables efficient auditing that does not require full data replication or
intrusive monitoring, thereby preserving privacy and remaining practical. We
formalize the protocol, analyze its security and performance properties, and
provide a reference implementation built on AWS infrastructure.  
\end{abstract}


\maketitle

\section{Introduction}
The rapid adoption of AI and advanced analytics has significantly increased the
demand for cross-organization data sharing, as high-performing models and
analytical systems depend on large, diverse datasets. At the same time, both
public- and private-sector organizations are navigating an increasingly
complicated data governance landscape. Regulatory frameworks such as the
California Privacy Rights Act (CPRA) and the General Data Protection Regulation
(GDPR), as well as e-discovery and litigation obligations, impose strict
restrictions on data sharing. But as data moves across organizational
boundaries, it is frequently transformed, filtered, or restructured in ways that
make linking shared data back to an original data resource difficult. Without
a record of what data is shared, from where, and with whom, assessing compliance
with these data sharing regulations can be challenging, or even impossible. 

To make these issues more concrete, consider two real-world scenarios. In
2023, the CPRA came into effect, giving consumers the right to opt out of the
sharing of personal information for marketing purposes~\cite{californiaagccpa}.
Soon thereafter, the US Congress considered (and the House of Representatives
passed) the Fourth Amendment Is Not for Sale Act, which would have forbidden law
enforcement agencies from acquiring Americans' data from data
brokers~\cite{congress2023fourthamendment}. Neither law, however, requires
regulated entities to preserve information about the content, origin, and
destination of shared data. Without this information, known generally as ``data
provenance,'' auditors lack the means to bring these laws to bear.

To vindicate the privacy goals of these and other laws, it's necessary to
develop a framework that makes clear how and what information must be preserved
to audit cross-organizational data sharing. Although many provenance frameworks
have been developed, there is no framework, to our knowledge, that addresses
this particular set of problems. Most of the prior work on provenance is
principally concerned with guaranteeing the
accuracy~\cite{opm2011, groth2013provoverview} and
security~\cite{hasan2009preventing,liang2017provchain, ruan2021lineagechain,
aldeco2010securing} of the provenance log. But maintaining a provenance log is
not on its own sufficient; certain transformations or migrations of the source
or destination databases can render the provenance log stale, thwarting
compliance auditing. Thus, a workable framework must define what additional
information must be retained to enable auditing. Moreover, the framework must
also supply a mechanism for auditors to verify that these requirements are being
met, while not forcing parties to divulge the content of potentially sensitive
data.

To fill this gap, we introduce \textit{Proof-of-Retention}, a novel framework
for ensuring auditable sharing of structured tabular data. We start by
formalizing data sharing as a contract between the sender and recipient. That
contract defines the minimum information necessary to be preserved to enable
auditing. In particular, we define each party's obligations around the goal of
avoiding \textit{provenance anomalies}, or, database states where recovering
provenance isn't possible. Provenance anomalies arise when either the sender
or destination unilaterally modifies their database, rendering the provenance
log stale. Importantly, the contract specifies \textit{what} must be preserved
rather than \textit{how} it must be preserved. Organizations then remain free to
satisfy the contract using any combination of database snapshots, transaction
logs, or other mechanisms specific to a given implementation. This separation
between obligation and implementation allows the contract to accommodate diverse
technical setups while providing a common standard for auditability.

From here, we define a sharing protocol to facilitate contract execution.
Parties exchange data via a trusted third-party \textit{broker} that is
responsible for transmitting the desired data and preserving a provenance
mapping between the sender and recipient databases. As discussed, a mapping
alone is insufficient to detect all provenance anomalies. Thus, an auditor must
be able to identify when a party updates or deletes data in violation of the
data exchange contract. To detect such violations, we define a verification
protocol that follows an interactive proof structure where the parties need not
reveal the retained data to the auditor to prove retention. First, the broker
generates and persists a (contract-specified) number of challenges at share time
by hashing the information required to be retained with random nonces. To verify
retention, the auditor then sends the nonce, and challenges the party to compute
the correct hash. We also design a reference implementation using commercially
available infrastructure for each of these protocols.

As data sharing across control boundaries becomes more common, the risk of private information being
leaked or misused grows. Entities that deploy AI agents---whether private firms or government
actors--for data analysis or data management tasks have both easier and more opaque access to
potentially sensitive data. Although existing privacy laws, including the CPRA and the proposed FANFSA,
seek to curb privacy violations arising from data sharing by forbidding or placing guardrails around
certain data. At the same time, existing methods for ensuring compliance with those laws are lacking
in two ways: (1) the obligations of regulated parties are either not defined or ill-defined and (2)
there is no way to ensure that those obligations are satisfied.

Our framework seeks to fill both gaps. We do not build the implementation, nor
do we conduct any experiments, because there is no baseline against which to
compare. Our contributions can be summarized as follows:
\begin{enumerate}[leftmargin=*, label=\arabic*.]
  \item \textbf{Formalization of Data Exchange} (Sec~\ref{sec:data-exchange}
    and~\ref{sec:contract}). We formalize
    a data exchange and each party's obligations to ensure the exchange is
    auditable around the use and retention of a witness.
  \item \textbf{Protocol for Auditing Retention}
    (Sec~\ref{sec:protocol}). We
    create a data exchange protocol that specifies how parties must share data
    to ensure auditability, as well as a verification protocol to ensure that
    parties retain the information necessary to audit each exchange.
  \item \textbf{Reference Implementation}
    (Sec~\ref{sec:threat-model}).
    We design a reference implementation for the broker and sharing protocol
    using commercial cloud infrastructure.
\end{enumerate}

\section{Background and Motivation}
In this section, we introduce concepts from the provenance literature that will be helpful to
understanding our framework, and motivate \textit{Proof-of-Retention}.

\subsection{Provenance}
In general, provenance~\cite{groth2013provoverview,opm2011} in structured databases relates the output tuples of a query $Q$ over a
database $\mathcal{D}$ to sets of tuples in $\mathcal{D}$. Several provenance schemes have been
developed to capture different relations between input and outputs as data is
transformed~\cite{cheney2009provenance}. In this work, we use a specific notion of provenance
referred to as \textit{why-provenance}~\cite{buneman2001and}. At a high level, the why-provenance of
a tuple $t$ in the output of $Q(\mathcal{D})$ is the family of input tuple sets that are sufficient
to produce $t$ according to $Q$. That family of sets is referred to as the \textit{witness basis} of
$t$, and each set a $witness$ of $t$~\cite{cheney2009provenance}.

\begin{figure}[h]
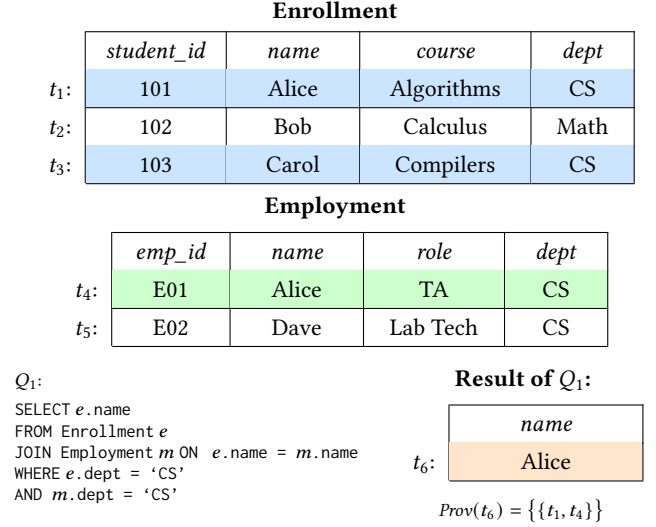

  \centering
  \renewcommand{\arraystretch}{1.3}

\centering
\textbf{Enrollment}\\[4pt]
\begin{tabular}{r|C{1.5cm}|C{1.4cm}|C{1.9cm}|C{1.0cm}|}
  \cline{2-5}
  & \textit{student\_id} & \textit{name} & \textit{course} & \textit{dept} \\
  \cline{2-5}
  $t_1$: & \cellcolor{enrollfill}101 & \cellcolor{enrollfill}Alice
          & \cellcolor{enrollfill}Algorithms & \cellcolor{enrollfill}CS \\
  \cline{2-5}
  $t_2$: & 102 & Bob & Calculus & Math \\
  \cline{2-5}
  $t_3$: & \cellcolor{enrollfill}103 & \cellcolor{enrollfill}Carol
          & \cellcolor{enrollfill}Compilers  & \cellcolor{enrollfill}CS \\
  \cline{2-5}
\end{tabular}

\medskip

\textbf{Employment}\\[4pt]
\begin{tabular}{r|C{1.2cm}|C{1.4cm}|C{1.5cm}|C{1.0cm}|}
  \cline{2-5}
  & \textit{emp\_id} & \textit{name} & \textit{role} & \textit{dept} \\
  \cline{2-5}
  $t_4$: & \cellcolor{empfill}E01 & \cellcolor{empfill}Alice
          & \cellcolor{empfill}TA & \cellcolor{empfill}CS \\
  \cline{2-5}
  $t_5$: & E02 & Dave & Lab Tech & CS \\
  \cline{2-5}
\end{tabular}

\bigskip
\noindent
\begin{minipage}[t]{0.55\linewidth}
  \footnotesize
  $Q_1$:\\[3pt]
  \texttt{SELECT} $e$\texttt{.name}\\
  \texttt{FROM Enrollment} $e$\\
  \texttt{JOIN Employment} $m$ \texttt{ON } $e$\texttt{.name = }$m$\texttt{.name}\\
  \texttt{WHERE} $e$\texttt{.dept = `CS'}\\
  \texttt{AND }$m$\texttt{.dept = `CS'}
\end{minipage}%
\hfill
\begin{minipage}[t]{0.40\linewidth}
  \centering
  \textbf{Result of $Q_1$:}\\[4pt]
  \begin{tabular}{r|C{2.2cm}|}
    \cline{2-2}
    & \textit{name} \\
    \cline{2-2}
    $t_6$: & \cellcolor{outfill}Alice \\
    \cline{2-2}
  \end{tabular}\\[6pt]
  \footnotesize
  $\textit{Prov}(t_6) = \bigl\{\{t_1,t_4\}\bigr\}$
\end{minipage}

  \caption{An example of why-provenance. Figure converted to LateX using AI}
  \label{fig:why-prov}
\end{figure}

Figure~\ref{fig:why-prov} illustrates the basic idea of why-provenance. Evaluating $Q_1$ over the
\texttt{Enrollment} and \texttt{Employment} relations produces a single output tuple $t_6$. The
why-provenance of $t_6$ includes the set of tuples $\{t_1, t_4\}$ because evaluating $Q_1$ over that
set alone is sufficient to produce $t_6$. Note that a witness for $t_6$ can include input tuples
that are not necessary to produce $t_6$ (e.g. $\{t_1, t_2, t_4\}$ also ``witnesses'' $t_6$). Indeed,
the degenerate witness for $t_6$ includes the entire input database. Thus, it can be useful to
distinguish \textit{minimal witnesses}, or those that only contain necessary
tuples~\cite{buneman2001and}.

From here, any reference to provenance in this paper refers to why-provenance. We choose
why-provenance because it is both simple and most relevant for tracing the movement of tuples that
are shared across organizational boundaries. Our framework could be extended to support other
notions of provenance, including \textit{how-provenance} (describing how an output tuple was derived
from a set of input tuples) and \textit{where-provenance} (mapping output and input cell locations).
But such extensions are left for future work.

Finally, provenance information can be queried. For a given tuple $t$, a
\textit{forward provenance query} discloses all tuples to which $t$ contributed. Conversely,
a \textit{backward provenance query} outputs the witness for a given tuple
$t$~\cite{kementsietsidis2009efficiency, kementsietsidis2009provenance}. Both types of provenance
queries elucidate the movement of data and make it possible to audit data sharing. But these queries
are only possible if there aren't errors in the provenance. In other words, there are no
provenance anomalies.

\subsection{The Problem of Provenance Anomalies}
Relational database theory has long studied anomalies in the context of normalization, where
redundancy can lead to inconsistencies during updates. For example, a relation containing a
transitive dependency may require multiple tuples to be modified to update a single logical fact,
creating the potential for inconsistent database states. Similarly, when data are exchanged across
data governance boundaries, organizational boundaries where different parties define and enforce
distinct data management systems, policies, and practices, new classes of anomalies can emerge.
These anomalies arise because data are replicated, transformed, and managed independently across
governance domains, creating additional opportunities for inconsistencies that extend beyond those
addressed by traditional normalization theory.

We are particularly interested in provenance anomalies, which arise when the
provenance associated with a data exchange event becomes stale because the
source and destination databases evolve under independent administrative
control. Unlike traditional update anomalies, which occur within a single
database due to redundancy, provenance anomalies occur across data governance
boundaries where neither party has complete authority to coordinate updates. As
the source database changes through insertions, deletions, or modifications, a
previously computed provenance relation may no longer accurately explain the
contents of the destination database. Conversely, independent updates to the
destination may invalidate assumptions about the source data on which the
provenance was originally based. As a result, provenance records that were once
correct can become incomplete, incorrect, or misleading, reducing their value
for auditing and regulatory compliance.

To better appreciate the problems of provenance anomalies, consider the following examples. Suppose
a marketing firm obtains user data from a data broker on the belief that all users have consented to
this data sharing, so as to comply with the CPRA. The firm further records the provenance of the
data exchange, so that a future auditor could validate that the users whose data was shared did
indeed consent. If, however, the sender migrates its database, leading to some user data being
deleted and other users being assigned new tuple ids, the provenance record will be stale, making it
difficult or impossible to connect the shared user data with the information disclosing whether the
user consented to the sharing.

In general, any changes to tuples in the witness of an output tuple can give rise to a provenance
anomaly. It is for this reason that merely updating the provenance log doesn't eliminate provenance
anomalies. Indeed, one way to conceptualize provenance is as evidence for a query output. If the
evidence is deleted or damaged, the provenance cannot be reconstituted. In the next section, we
formalize the concepts and intuitions presented so far.

\section{Data Exchange and Witnesses}\label{sec:data-exchange}
In this section, we formalize a data exchange, provenance, and witness, as well as introduce the
notation to be used throughout the rest of the paper.

\subsection{Data Exchange}
Data exchange has been extensively studied in database theory, particularly in
the context of data integration, data quality, missing data, and entity
resolution~\cite{fagin2005data}. In this work, we consider data exchanges that
happen across \emph{data governance boundaries} and try to address the
provenance challenges that arise when independently administered databases
evolve after an exchange event. 

We use a standard relational model where a source agent shares data with a destination agent. Each
relation symbol $R \in \mathbf{S}$ has an ordered list of attributes $\mathrm{att}(R)$, a finite
arity, and a primary-key attribute set $\mathrm{key}(R) \subseteq \mathrm{att}(R)$. An
\emph{instance} $I$ of $\mathbf{S}$ maps each $R \in \mathbf{S}$ to a finite relation $R^I$ over the
attributes $\mathrm{att}(R)$ that satisfies the key and any other constraints of $\mathbf{S}$.

Primary keys give every stored tuple a stable logical name. For a relation $R_i \in \mathbf{S}$ and
tuple $t \in R_i^I$, let $\mathrm{tid}_{src}(t)$ denote the pair consisting of $R_i$ and the value
of $\mathrm{key}(R_i)$ in $t$ in the source database. The destination has analogous identifiers
$\mathrm{tid}_{dst}(u)$ for tuples $u \in T_j^J$. These tuple identifiers are logical identifiers,
not assumptions about the physical layout of either database.

\vspace{0.25em} \noindent \textbf{An Exchange Event. }
An exchange event involves two database instances over two generally different schemas. The
\emph{source database} is
\[
  \mathcal{D}_{src}=(\mathbf{S}, I),
\]
where $\mathbf{S}$ is the source schema and $I$ is the source instance. The \emph{destination
database} is
\[
  \mathcal{D}_{dst}=(\mathbf{T}, J),
\]
where $\mathbf{T}$ is the destination schema and $J$ is the destination instance before the sharing
event. We reserve $S$ for the source actor and $D$ for the destination actor; relation symbols are
written as $R_i,T_j$ when needed.

A data exchange event is modeled as a data exchange from the source schema to the destination
schema. A \emph{data exchange formula}, denoted by $\Delta$, is a deterministic program that takes
as input $\mathcal{D}_{src}$ to produce new tuples that are inserted into $\mathcal{D}_{dst}$. The
particular language used to express $\Delta$ is not important for our development; we assume only
that it is deterministic. Throughout the paper we focus on relational mappings. That is, $\Delta$
consists of a collection of relational queries that apply selection, projection, joins, aggregation,
and renaming to produce tuples that are to be inserted into $\mathcal{D}_{dst}$.

Applying $\Delta$ to $\mathcal{D}_{src}$ and inserting the resulting tuples into $\mathcal{D}_{dst}$
produces a post-exchange destination instance $J'$. Throughout this paper we assume insert-only data
exchange: tuples produced by $\Delta$ are inserted into the destination database only if they do not
already exist. Updates and deletions are outside the scope of this work.

\subsection{Provenance in Exchange Events}\label{subsec:prov-def}
Every exchange event is accompanied by a provenance relation that records how source tuples
contribute to newly inserted destination tuples. Thus, for a formula $\Delta$, $I$ (source
instance), and $J'$ (post-exchange destination instance), it is possible to calculate a mapping from
tuple IDs in $I$ to those newly inserted in $J'$:
\[ \prov(\Delta, I, J') \subseteq \mathrm{tid}_{src}(I) \times \mathrm{tid}_{dst}(J'), \]
which records the tuple-level correspondence
induced by the exchange. That is, it describes some relation over input and output tuples. Note that
this is a definition of why-provenance, but can be extended to other provenance definitions as well.
We defer that discussion to future work. Over this relation, a \emph{forward provenance query}
identifies all tuples in $J'$ related to a particular tuple in $I$. Similarly, a \emph{backward
provenance query} identifies all tuples in $I$ that are related to a particular tuple in $J'$

The provenance relation cannot be an arbitrary correspondence between tuples; rather, it must
faithfully describe how the exchange process produced the post-exchange database. So, the definition
of $\prov(\Delta, I, J')$ is related to the concept of a witness in database theory. In database
theory, a witness is a subdatabase that is sufficient to demonstrate why a query result is present
or why a particular property holds. Witnesses are a fundamental concept in provenance, causality,
and query evaluation because they provide a concise explanation that links an output directly to the
input data responsible for producing it.

Projecting the provenance relation onto its source tuple identifiers naturally defines a subset of
the source database. The corresponding subdatabase is denoted by $W_{\Delta}$ by only
selecting those tuples in $\mathcal{D}_{src}$. Provenance relations that can generate such
\emph{valid} witnesses are called witness-generating provenance (WGP):

\begin{definition}[Witness-Generating Provenance]
Let $\prov(\Delta,I,J')$ denote the provenance relation for a data exchange
event. We say that $\prov(\Delta,I,J')$ is \emph{witness-generating} if it
identifies a witness $W_{\Delta}\subseteq\mathcal{D}_{src}$ consisting
of the source tuples referenced by the provenance relation, and executing
$\Delta$ over $W_{\Delta}$ produces exactly the same post-exchange
destination instance $J'$.
\end{definition}

A witness-generating provenance relation captures not only the correspondence
between source and destination tuples but also identifies a subset of the source
database that is sufficient to reproduce the exchanged data. By construction,
executing the exchange formula over $W_{\Delta}$ produces exactly the
same post-exchange instance $J'$. Consequently, every tuple outside
$W_{\Delta}$ is irrelevant to the outcome of that particular exchange
event.

\begin{definition}[Valid Witness]\label{def:valid-witness}
A \emph{valid} witness for a data exchange event $(\Delta, I, J')$ is on that
exactly reproduces $J'$ if $\Delta$ is applied to it.
\end{definition}

\subsection{Satisfying Witnesses}
Several observations follow from this definition. First, the witness provides a constructive
explanation of the data exchange, making it useful for provenance queries, auditing, and debugging.
Second, requiring that $W_{\Delta}$ alone reproduce $J'$ distinguishes witness-generating
provenance from provenance representations that merely record dependencies without guaranteeing
sufficiency. Finally, the definition requires only that the witness be sufficient—not unique or
globally minimal. Multiple witnesses may satisfy the definition depending on the semantics of the
mapping.

A careful reader will further note that the definition of a witness is underspecified in terms of
attributes, meaning that different projections of the relevant tuples can yield valid witnesses as
long as they produce $J'$. We model this requirement by assuming a designated set of required
attributes. Formally, let $P \subseteq \mathrm{att}(R)$. 

\begin{definition}[Satisfying Witness]\label{def:satisfying-witness}
A \emph{satisfying} witness for a data exchange event $(\Delta, I, J')$ is a
witness such that $P \subseteq \mathrm{att}(W_{\Delta})$.
\end{definition}

\subsection{Provenance Anomalies in Exchanges}

With the preceding formalism, we can precisely characterize the concept of a provenance anomaly for
an exchange event.

\begin{definition}[Provenance Anomaly]
  A provenance anomaly occurs when a source-destination pair is unable to produce a verifiably valid
  and satisfying witness for a given exchange event $(\Delta, I, J')$.
\end{definition}

It might be easier to think about this definition in the converse. Answering provenance queries
accurately requires the source-destination pair to produce a witness for any past data exchange event
and present evidence that the witness is valid and satisfying. The fault of a failure in this
condition may lie on either side. For example, if the source did not retain sufficient data to
produce a witness, that failure is an anomaly because backward queries can no longer be answered accurately. Alternatively,
if the destination failed to retain the new tuples added in $J'$, then forward queries cannot be
accurately answered.

\section{Auditable Data Exchange}\label{sec:contract}
The existence of provenance anomalies motivates a stronger correctness property for data exchange
systems. Rather than simply recording provenance at the time of an exchange, we require that
provenance remain sufficient to independently verify the exchange after the participating databases
have evolved. An auditable data exchange system guarantees that every exchange event can be
justified by a witness that is both valid, in that it corresponds to the current source data, and
satisfying, in that it contains all information required to reproduce the exchanged destination
data. Consequently, every exchange remains independently verifiable, even when the sender and
destination are administered under separate data governance policies.

\subsection{A Data Exchange Contract}
We formalize these obligations as an \emph{auditable data exchange contract}. Rather than
prescribing a particular implementation, the contract specifies the responsibilities of each party
that are necessary to ensure that every completed exchange remains independently verifiable. In
particular, the contract guarantees that every exchange event admits a witness that is both valid
and satisfying.

An \emph{auditable data exchange contract} between a sender $S$ and destination $R$ for an exchange
mapping $\Delta$ is an agreement satisfying the following properties for every completed exchange
event.

\begin{enumerate}
    \item \textbf{Witness Preservation.}
    The sender maintains sufficient information to construct a valid witness
    $W_{\Delta}$ for the exchange event.

    \item \textbf{Destination Preservation.}
    The destination maintains sufficient information to identify the exchanged
    destination tuples associated with the post-exchange instance $J'$. Specifically,
    in an insertion-only model this means any tuples $t \in J'$ but not in $J$.

    \item \textbf{Witness Production.}
    Upon request, the sender and destination can jointly produce a witness and destination tuples.

    \item \textbf{Witness Verification.}
    Either party, or an independent auditor, can verify that the produced
    witness is valid and satisfying according to
    Definitions~\ref{def:valid-witness}
    and~\ref{def:satisfying-witness}.
\end{enumerate}

The contract intentionally abstracts away implementation details. It does not
require either party to retain complete historical snapshots of their databases,
nor does it prescribe how provenance or auxiliary metadata are stored. Instead,
it merely specifies what information must be preserved to audit the data
exchange. As such, parties are free to use a particular implementation, so long
as it enables the production of a valid and satisfying witness for each
exchange. Formally: 

\begin{definition}[Auditable Data Exchange Contract]\label{def:contract}
An auditable data exchange contract is a tuple
$K=(\mathcal O_S,\mathcal O_D,\mathcal V)$\footnote{$\mathcal{K}$ is the common legal shorthand for
a contract} where

\begin{itemize}
\item $\mathcal O_S$ specifies the information that the sender must preserve for every exchange event;
\item $\mathcal O_D$ specifies the information that the destination must preserve;
\item $\mathcal V$ is a deterministic verification procedure that accepts a witness iff it is valid and satisfying.
\end{itemize}

The contract is satisfied if, for every completed exchange event, the obligations
$\mathcal O_S$ and $\mathcal O_D$ are sufficient for $\mathcal V$ to verify a
valid and satisfying witness.
\end{definition}

By tying every exchange to a valid and satisfying witness, the contract
guarantees that provenance is either (a) preserved across organizational
boundaries or (b) it is possible to detect provenance anomalies. This property
is particularly valuable in settings where data routinely moves between
independently governed systems, such as government agencies, healthcare
providers, and financial institutions. An auditable data exchange contract
ensures that provenance relationships remain verifiable even as the databases
involved in the exchange continue to evolve independently.

Finally, the contract defines an objectively verifiable notion of compliance.
Rather than relying on policy statements or organizational trust, performance of
the contract can be evaluated by attempting to produce and verify valid and
satisfying witnesses for historical exchange events. Moreover, regulators can
now supply and require the use of ``boilerplate'' contracts for entities in
specific industries or when specific data is exchanged. That way, regulators can
say with specificity what is required, while making it easier to verify
compliance. Indeed, compliance is judged immediately based on whether the
parties can produce a valid and satisfying witness. Consequently, auditability
becomes a property of a data exchange relationship rather than an aspirational
organizational goal, all while providing a rigorous foundation for regulatory
compliance and auditing.

\subsection{Further Witness Constraints}
So far, a witness is satisfying so long as $P \subseteq \mathrm{att}(W_{\Delta})$, where
$P$ is the set of required attributes. But the data exchange contract is extensible, allowing for
additional constraints to be required of the witness. To start, the contract might require that
the witness tuples adhere to a specific format (e.g., MM/DD/YYYY rather than DD/MM/YYYY).

For certain data exchange events, the contract may require additional attributes, or other
identifiers not necessary to produce the same post-exchange instance $J'$. If, for example, the data
exchange event acts over anonymized data (i.e. lacking a particular identifier), the contract could
optionally require the sender to retain a witness that preserves a link between the shared data and
a particular identifier. We refer to such requirements as \textit{linkage constraints}. Two examples
of linkage constraints follow:

\subsubsection{Tuple-Level Linkage}. For some $(\Delta, I, J')$, a tuple-level linkage constraint
requires that, for each tuple $t \in J'$ but not in $J$, the witness contains a functional dependency between
$t$ and some chosen identifier attribute $Y \subseteq \mathrm{att}(\mathrm{S})$. For example, the
source might share anonymized patient data with the destination to analyze. If the destination's
analysis uncovers information that must be relayed to a particular patient, the destination might
require the source to preserve a mapping between the shared anonymous patient ID and some patient identifier. 

\subsubsection{Set-Level Linkage}. For some $(\Delta, I, J')$ where $\Delta$ applies an aggregation,
a set-level constraint requires that the witness contain enough information to recover the
identifiers $Y$ of the source tuples contributing to each group. For example, an auditor may
determine that an advertising firm has created an aggreggate metric using many users' data, in
violation of the CPRA. To be able to determine which users were involved, the contract would need to
specify a set-level constraint between the users and the aggregate metric. 

\subsection{Executing the Contract}
Having defined provenance anomalies and introduced the data exchange contract primitive, we can
elaborate the desiderata for how the contract must be executed. In other words, how we can ensure
that both parties' contract obligations are met while not compromising privacy or practicality.

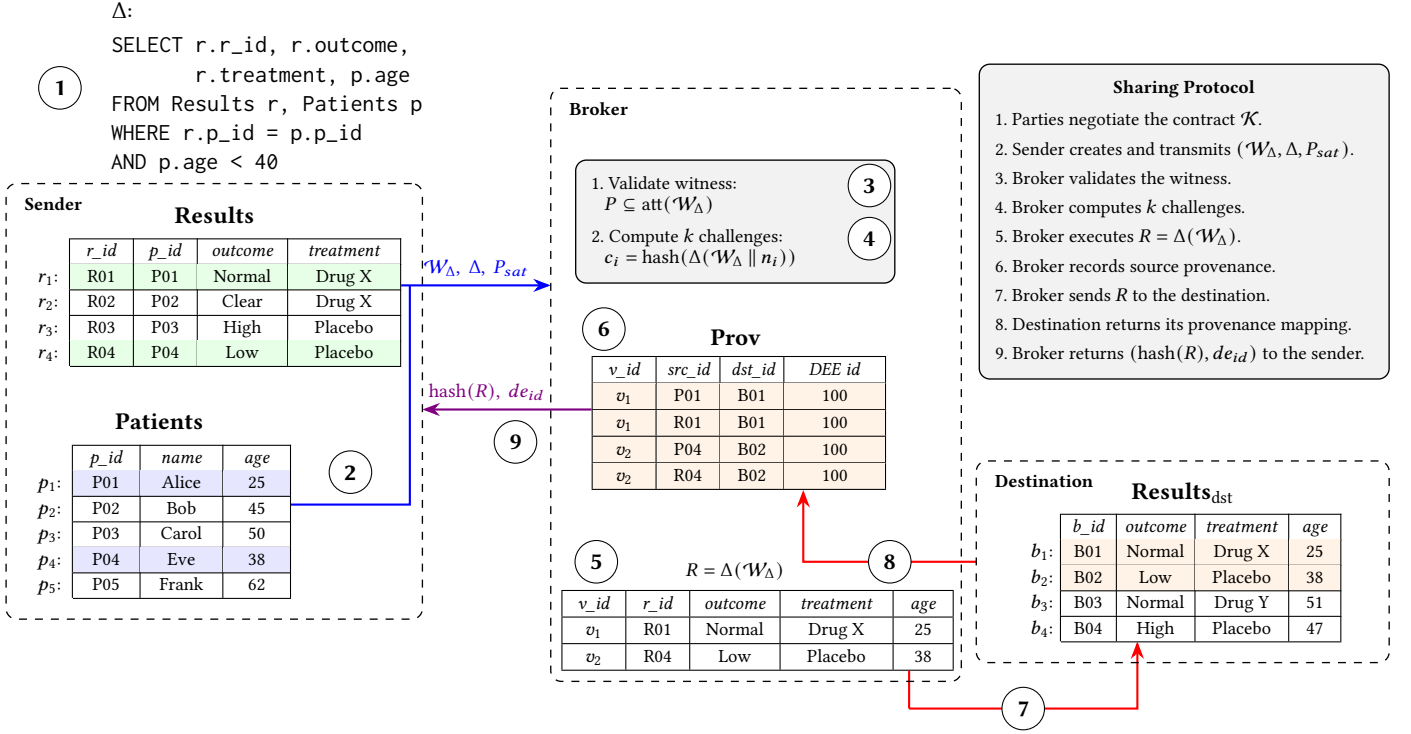
\begin{figure*}[!t]
  \centering
  \renewcommand{\arraystretch}{1.2}
\setlength{\tabcolsep}{4pt}
\usetikzlibrary{calc, fit}
\begin{tikzpicture}[>=Stealth, line width=0.5pt, xshift=-1cm,
    ann/.style={font=\footnotesize\sffamily},
    tbl/.style={inner sep=0pt, outer sep=0pt},
    circnum/.style={circle, draw=black, solid, thin, inner sep=1.5pt, minimum size=16pt, font=\normalsize\sffamily\bfseries, text=black, fill=white}]
\colorlet{pfill}{blue!10}
\colorlet{rfill}{green!10}
\colorlet{vfill}{orange!10}

%% ── Left column: Results (top), Patients (below) ─────────────────────────────
%% Both tables are right-aligned at x = -4.2 so their east edges line up.
\node[tbl, anchor=north east] (results) at (-16.5,0) {\footnotesize
  \begin{tabular}{r|C{0.55cm}|C{0.55cm}|C{0.9cm}|C{1.2cm}|}
    \multicolumn{5}{c}{\normalsize\textbf{Results}}\\[3pt]
    \cline{2-5}
    & \textit{r\_id} & \textit{p\_id} & \textit{outcome} & \textit{treatment} \\
    \cline{2-5}
    $r_1$: & \cellcolor{rfill}R01 & \cellcolor{rfill}P01 & \cellcolor{rfill}Normal & \cellcolor{rfill}Drug X \\
    \cline{2-5}
    $r_2$: & R02 & P02 & Clear & Drug X \\
    \cline{2-5}
    $r_3$: & R03 & P03 & High & Placebo \\
    \cline{2-5}
    $r_4$: & \cellcolor{rfill}R04 & \cellcolor{rfill}P04 & \cellcolor{rfill}Low & \cellcolor{rfill}Placebo \\
  \cline{2-5}
  \end{tabular}%
};

% Anchor Patients so its east edge also sits at x = -4.2, 0.6cm below Results
\path (results.south west) ++(0,-0.6) coordinate (patAnchor);
\node[tbl, anchor=north west] (patients) at (patAnchor) {\footnotesize
  \begin{tabular}{r|C{0.6cm}|C{0.8cm}|C{0.6cm}|}
    \multicolumn{4}{c}{\normalsize\textbf{Patients}}\\[3pt]
    \cline{2-4}
    & \textit{p\_id} & \textit{name} & \textit{age} \\
    \cline{2-4}
    $p_1$: & \cellcolor{pfill}P01 & \cellcolor{pfill}Alice & \cellcolor{pfill}25 \\
    \cline{2-4}
    $p_2$: & P02 & Bob & 45 \\
    \cline{2-4}
    $p_3$: & P03 & Carol & 50 \\
    \cline{2-4}
    $p_4$: & \cellcolor{pfill}P04 & \cellcolor{pfill}Eve & \cellcolor{pfill}38 \\
    \cline{2-4}
    $p_5$: & P05 & Frank & 62 \\
    \cline{2-4}
  \end{tabular}%
};

%% ── Top: Queries ─────────────────────────────────────────────────────────────
\node[anchor=south west, font=\normalsize\sffamily] (queries) at ([xshift=1.0cm, yshift=0.3cm]results.north west) {%
  \begin{minipage}[t]{5.5cm}
    $\Delta$:\\[2pt]
    \texttt{SELECT r.r\_id, r.outcome,}\\
    \texttt{\phantom{SELECT }r.treatment, p.age}\\
    \texttt{FROM Results r, Patients p}\\
    \texttt{WHERE r.p\_id = p.p\_id}\\
    \texttt{AND p.age < 40}
  \end{minipage}%
};
% Circle 1: Placed in front of the queries
\node[circnum, anchor=east] at ([xshift=-8pt]queries.west) {1};

%% ── Middle column: Steps (top), A-to-B (middle), V(Q) (bottom)
\node[anchor=north west, draw, rounded corners, fill=gray!10, inner sep=6pt] (steps) at (-14.2,0.6) {%
  \begin{minipage}{3.8cm}
    \footnotesize
    1. Validate witness:\\
    \phantom{1.}$P \subseteq \mathrm{att}(\mathcal{W}_{\Delta})$\\[4pt]
    2. Compute $k$ challenges:\\
    \phantom{2.}$c_i = \mathrm{hash}(\Delta(\mathcal{W}_{\Delta} \,\|\, n_i))$
  \end{minipage}%
};
% Steps 3 and 4: witness validation and challenge creation.
\node[circnum, anchor=north east] at ([xshift=-4pt, yshift=-4pt]steps.north east) {3};
\node[circnum, anchor=north east] at ([xshift=-4pt, yshift=-24pt]steps.north east) {4};
\node[tbl, anchor=north, below=0.6cm of steps] (atob) {%
  \begin{minipage}{3.8cm}
    \centering\footnotesize
    {\normalsize\textbf{Prov}}\\[3pt]
    \begin{tabular}{|C{0.55cm}|C{0.55cm}|C{0.55cm}|C{1.05cm}|}
      \hline
      \textit{v\_id} & \textit{src\_id} & \textit{dst\_id} & \textit{DEE id} \\
      \hline
      \cellcolor{vfill}$v_1$ & \cellcolor{vfill}P01 & \cellcolor{vfill}B01 & \cellcolor{vfill}100 \\
      \hline
      \cellcolor{vfill}$v_1$ & \cellcolor{vfill}R01 & \cellcolor{vfill}B01 & \cellcolor{vfill}100 \\
      \hline
      \cellcolor{vfill}$v_2$ & \cellcolor{vfill}P04 & \cellcolor{vfill}B02 & \cellcolor{vfill}100 \\
      \hline
      \cellcolor{vfill}$v_2$ & \cellcolor{vfill}R04 & \cellcolor{vfill}B02 & \cellcolor{vfill}100 \\
      \hline
    \end{tabular}
  \end{minipage}%
};
% Circle 6: Broker records provenance to the ledger.
\node[circnum, anchor=east] (circ6) at ([xshift=-1.45cm]atob.north) {6};
\node[tbl, anchor=north, below=0.95cm of atob] (vq) {%
  \begin{minipage}{4.6cm}
    \centering\footnotesize
    \textbf{$R = \Delta(\mathcal{W}_{\Delta})$}\\[3pt]
    \begin{tabular}{|C{0.55cm}|C{0.55cm}|C{0.9cm}|C{1.2cm}|C{0.5cm}|}
      \hline
      \textit{v\_id} & \textit{r\_id} & \textit{outcome} & \textit{treatment} & \textit{age} \\
      \hline
      $v_1$ & R01 & Normal & Drug X & 25 \\
      \hline
      $v_2$ & R04 & Low & Placebo & 38 \\
      \hline
    \end{tabular}
  \end{minipage}%
};

%% ── Top Right: Sharing Protocol List (Moved to Top Right) ───────────────────
\node[anchor=north west, draw, rounded corners, fill=gray!10, inner sep=6pt, right=1.1cm of steps] (newsteps) {%
  \begin{minipage}{5.0cm}
    {\centering\footnotesize\textbf{Sharing Protocol}\par\vspace{4pt}}
    \footnotesize
    1. Parties negotiate the contract $\mathcal{K}$.\\[3pt]
    2. Sender creates and transmits $(\mathcal{W}_{\Delta},\Delta,P_{sat})$.\\[3pt]
    3. Broker validates the witness.\\[3pt]
    4. Broker computes $k$ challenges.\\[3pt]
    5. Broker executes $R=\Delta(\mathcal{W}_{\Delta})$.\\[3pt]
    6. Broker records source provenance.\\[3pt]
    7. Broker sends $R$ to the destination.\\[3pt]
    8. Destination returns its provenance mapping.\\[3pt]
    9. Broker returns $(\mathrm{hash}(R),de_{id})$ to the sender.
  \end{minipage}%
};

%% ── Right column: Results table (Moved Down) ────────────────────────────────
% The Results table is vertically centered between A-to-B and V(Q)
\path (atob.south) -- (vq.north) coordinate (midPt);
\node[tbl, anchor=center] (bresults) at ([xshift=3.8cm]steps.east |- midPt) {%
  \begin{minipage}{4.9cm}
    \centering\footnotesize
    {\normalsize\textbf{Results}$_{\text{dst}}$}\\[3pt]
    {\setlength{\tabcolsep}{2.5pt}
    \begin{tabular}{r|C{0.55cm}|C{0.85cm}|C{1.05cm}|C{0.5cm}|}
      \cline{2-5}
      & \textit{b\_id} & \textit{outcome} & \textit{treatment} & \textit{age} \\
      \cline{2-5}
      $b_1$: & \cellcolor{vfill}B01 & \cellcolor{vfill}Normal & \cellcolor{vfill}Drug X & \cellcolor{vfill}25 \\
      \cline{2-5}
      $b_2$: & \cellcolor{vfill}B02 & \cellcolor{vfill}Low & \cellcolor{vfill}Placebo & \cellcolor{vfill}38 \\
      \cline{2-5}
      $b_3$: & B03 & Normal & Drug Y & 51 \\
      \cline{2-5}
      $b_4$: & B04 & High & Placebo & 47 \\
      \cline{2-5}
    \end{tabular}}
  \end{minipage}%
};

%% ══════════════════════════════════════════════════════════════════════════════
%% ── Arrows ───────────────────────────────────────────────────────────────────
%% ══════════════════════════════════════════════════════════════════════════════

%% ---------- Set 1 (BLUE): Results + Patients  -->  Steps ---------
\coordinate (blueR) at (results.east);
\coordinate (blueEnd) at ([xshift=-4pt]vq.west |- blueR);
\coordinate (blueTrunkX) at ($(blueR)!0.05!(steps.west)$);
% Main straight arrow from Results to Steps
\draw[->, blue, thick] (blueR) -- (blueEnd)
node[midway, above, ann, text=blue] {$\mathcal{W}_{\Delta},\ \Delta,\ P_{sat}$};
% Tap from Patients connects to the straight line
\coordinate (blueP) at (patients.east);
\draw[blue, thick] (blueP) -- (blueP -| blueTrunkX)
  node[midway, above=4pt, circnum] {2} -- (blueTrunkX |- blueR);

\node[circnum, anchor=east] at ([xshift=-1.55cm]vq.north) {5};

%% ---------- Set 3 (RED): V(Q) -> Results & Results -> A-to-B -----------
% Red line continues directly out of the Q table (connects to the "Q" arrow feeding vq.west),
% drops into clear space below vq, then turns right and up into the Destination box.
\coordinate (vqStart) at (vq.south east);
\coordinate (vqDrop) at ([yshift=-0.5cm]vqStart);
\coordinate (vqTurn) at ([xshift=-0.6cm]bresults.south);
\draw[red, thick] (vqStart) -- (vqDrop);
\draw[->, red, thick] (vqDrop) -- (vqDrop -| vqTurn) node[midway, circnum] {7} -- (vqTurn);

% Red line from rightmost Results table leftwards then up into A-to-B, entering from below
\coordinate (atobBelow) at ([xshift=-1.0cm]atob.south east);
\coordinate (stepEightStart) at ([xshift=-8pt]bresults.west);
\draw[->, red, thick] (stepEightStart) -- (stepEightStart -| atobBelow)
  node[midway, circnum] {8} -- (atobBelow);

%% ---------- Set 4 (PURPLE): Broker receipt -> Sender -----------------------
\coordinate (receiptStart) at (atob.west);
\coordinate (receiptEnd) at ([xshift=8pt]results.east |- receiptStart);
\draw[->, violet, thick] (receiptStart) -- (receiptEnd)
  node[pos=0.62, above, ann, text=violet] {$\mathrm{hash}(R),\ de_{id}$}
  node[pos=0.45, below=4pt, circnum] {9};

%% ══════════════════════════════════════════════════════════════════════════════
%% ── Governance boxes: who controls what ─────────────────────────────────────
%% ══════════════════════════════════════════════════════════════════════════════
\node[draw, dashed, rounded corners, inner sep=8pt, fit=(results)(patients)] (senderBox) {};
\node[anchor=north west, font=\bfseries\footnotesize, xshift=4pt, yshift=-2pt] at (senderBox.north west) {Sender};

\coordinate (brokerTop) at ([yshift=0.8cm]steps.north);
\coordinate (brokerRight) at ([xshift=0.95cm]atob.east);
\node[draw, dashed, rounded corners, inner sep=4pt, fit=(brokerTop)(brokerRight)(steps)(atob)(vq)(circ6)] (brokerBox) {};
\node[anchor=north west, font=\bfseries\footnotesize, xshift=4pt, yshift=-2pt] at (brokerBox.north west) {Broker};

\node[draw, dashed, rounded corners, inner sep=8pt, fit=(bresults)] (destBox) {};
\node[anchor=north west, font=\bfseries\footnotesize, xshift=4pt, yshift=-2pt] at (destBox.north west) {Destination};

\end{tikzpicture}
  \caption{A walk-through of the sharing protocol. Figure converted to LateX using AI}
  \label{fig:sharing-protocol}
\end{figure*}

Although Definition~\ref{def:contract} provides a precise definition of auditability, the preceding
section leaves open an important practical challenge. Demonstrating compliance with an auditable
data exchange contract appears to require the sender and destination to reveal the data whose
retention is being verified. In the simplest approach, the sender wouold present the witness
corresponding to a given data exchange, the destination reveals the exchanged tuples, and an auditor
could simply verify that the witness was both valid and satisfying. While correct, this approach is
impractical because the exchanged data may be confidential, regulated, or prohibitively large. Thus,
presenting the witness and exchanged data as is may violate privacy or governance constraints. 

In light of these constraints, we can formally articulate the following goals
for a framework to execute data exchange contracts:

\begin{goal}[Support Forward and Backwards Provenance Queries]\label{goal:prov-queries} 
  The framework should enable an auditor to answer forward and backward
  provenance queries, as provenance is defined in Section~\ref{subsec:prov-def}.
\end{goal}

Support for provenance queries is a strict requirement. It requires that (a) the
provenance of every data exchange is preserved and (b) there is a mechanism to
query that provenance information. Both sub-requirements are necessary to
prevent and audit provenance anomalies. For (a), not capturing provenance
information is itself a provenance anomaly (recall gaps in the provenance record
are types of anomalies). The ability to query provenance is in turn necessary to
audit data exchanges. Consider again our CPRA example, where an auditor wishes
to determine whether specific user data has been shared contrary to user
preferences. To answer this question, the auditor must run forward provenance
queries over that user's data. Alternatively, that auditor might target a
particular firm, checking whether the users of data they receive have consented
to the data exchange. Now, the auditor will perform backward provenance queries
over the firm's data.

\begin{goal}[Prove Compliance]\label{goal:compliance}  
  The framework must enable honest parties that have retained the information
  specified by the contract to be able to demonstrate compliance. Accordingly,
  parties that have not retained the necessary information should not be able to
  convince an auditor otherwise.
\end{goal}

\begin{goal}[Privacy-Preserving Verification]\label{goal:privacy-preserving} 
  The framework should provide a mechanism for the parties to prove compliance
  without revealing additional information about the data exchanged. 
\end{goal}

Proving compliance in a privacy-preserving manner is not, strictly speaking,
necessary to achieve our overall goal. Indeed, a trivial framework would require
parties to retain and make public every version of their database and provenance
record. But such a framework is practically undesirable for these two reasons.
Starting with privacy-preserving verifiability, the retained data may be
sensitive or contain PII, and not meant to be disclosed to even the auditor. In
such cases, it becomes necessary to use a protocol that proves data retention,
without revealing the contents of the data. 

Instead of revealing witnesses or destination tuples directly, the sender and
destination produce cryptographic evidence attesting to the continued retention
of the information required by the contract. An auditor can verify this evidence
to establish that a valid and satisfying witness could be reconstructed if
necessary, without learning the witness itself or the exchanged data. In this
sense, the protocol proves the ability to satisfy the contract rather than
reconstructing the underlying exchange.

We design \textit{Proof-of-Retention} around these three goals.
Proof-of-Retention is a framework that enables a sender and destination to
convince a verifier or auditor that they continue to possess sufficient information to
construct and verify a valid and satisfying witness for a historical exchange
event, all without revealing the witness or the exchanged data. The framework is
composed of two protocols, a sharing protocol
(Section~\ref{subsec:sharing-protocol} and a
verification protocol
(Section~\ref{subsec:verification-protocol}).
The rest of the paper walks through each of these protocols, as well as a
reference implementation for the framework.

\section{Sharing and Auditing Protocol}\label{sec:protocol}
This section begins by describing the sharing and verification protocols and then the broker implementation
needed to support each protocol. The sharing protocol defines the steps the \texttt{sender},
\texttt{broker}, and \texttt{destination} must take to share data such that it is possible to verify that
the necessary information about the data exchange has been retained.

\subsection{Sharing Protocol}\label{subsec:sharing-protocol}
The sharing protocol defines a procedure to execute an auditable data exchange contract
$\mathcal{K}$ (see Definition~\ref{def:contract}) over structured data between two parties. We
present the sharing protocol using the example source and destination databases in
Figure~\ref{fig:sharing-protocol}. In our example, the source
database $\mathcal{D}_{src}$ contains two relations, Results and Patients, storing test outcomes
keyed by the patient to which they belong and patient identifying information, respectively:
\[
  \begin{gathered}
    \textsf{Results}(\underline{\texttt{r\_id}}, \texttt{p\_id}, \texttt{outcome}, \texttt{treatment}) \\
    \textsf{Patients}(\underline{\texttt{p\_id}}, \texttt{name}, \texttt{age})
  \end{gathered}
\]
The destination database $\mathcal{D}_{dst}$ contains its own Results$_{dst}$
relation with the following schema: \[
  \textsf{Results}_{\text{dst}}(\underline{\texttt{b\_id}}, \texttt{outcome},
  \texttt{treatment}, \texttt{age})
\]
We now walk through each step of the sharing protocol. The broker implementation is described in
greater detail in Subsection~\ref{subsec:broker}. For now it suffices to treat the broker as a
trusted third-party service.\\

\noindent\textbf{Step 1: Data Exchange Contract Negotiation}. Before sharing, the \texttt{sender}
and \texttt{destination} must agree on each party's information preservation obligations, i.e., the
parameters $\mathcal{O}_{S}$ and $\mathcal{O}_{R}$ of the contract $\mathcal{K}$. This process will
typically happen out of band, with $\mathcal{O}_{S}$ defined in terms of the data to be exchanged.
For structured data, our protocol expects the parties to define the shared data in terms of a query
$\Delta$ executed over $\mathcal{D}_{src}$. From there, $\mathcal{O}_{src}$ is a valid witness
$W_{\Delta}$ such that $\Delta(W_{\Delta})=\Delta(\mathcal{D}_{src})$. In our example, $\Delta$ is a
query to retrieve anonymized test results for patients under 40 in the \texttt{sender}'s database.
For structured data, $\mathcal{O}_{R}$ is typically just the result $R= \Delta(W_{\Delta})$.

The parties may optionally agree to $P_{sat}$, a set of additional satisfiability requirements on
$W_{\Delta}$. In our example, the \texttt{destination} might require the \texttt{sender} to include the
\texttt{name} columns from the \texttt{Patients} relation in $W_{\Delta}$, even though that column
isn't necessary to produce the desired result. Such a requirement could be desirable if needed to
identify the patients of certain results, or, more generally, whenever the origin or context of
shared data is relevant to determining the appropriateness of the data exchange.

Finally, the parties must agree to a verification protocol $\mathcal{V}$. We propose one such
$\mathcal{V}$ in Subsection~\ref{subsec:verification-protocol} (parameterized by the number of
audits, $k$, and the frequency of audits), but the parties may supply their own $\mathcal{V}$.

\noindent\textbf{Step 2: Witness Creation and Transmission}. Once the parties have defined the data exchange
contract, the sender transmits $(W_{\Delta}, \Delta, P_{sat})$ to the broker.

\noindent\textbf{Step 3: Witness Validation}. Upon receipt, the broker must validate that the
witness is satisfying. The broker can verify satisfiability by checking that $P_{sat} \subseteq
\mathrm{att}(W_{\Delta})$.

\noindent\textbf{Step 4: Challenge Creation}. Once the witness has been validated, the broker must
compute a set of challenges for the sender. The challenges are used during the Verification Protocol
($\mathcal{V}$) to ensure that the sender has retained the data specified by $\mathcal{O}_{S}$. At
this stage, the broker must generate $k$ nonces, $n_i$, where $k$ is specified in the contract. The
broker then computes a set of $k$ challenges using a collision-resistant hash function $\texttt{h}:
\{0, 1\}^l \rightarrow \{0,1\}^{\lambda}$, $c_i = \texttt{h}(\Delta(W_{\Delta} || n_i))$ for $k$
nonces. The broker must then associate the set of $c_i$'s with this particular data sharing
contract, and each $c_i$ with the nonce $n_i$.

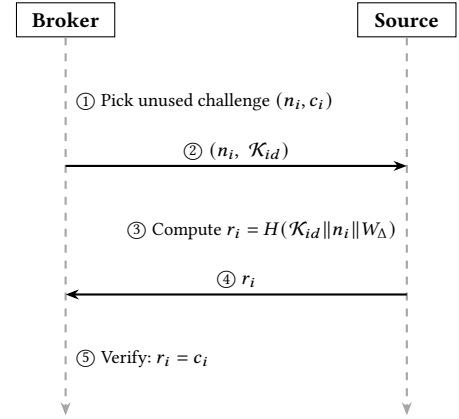
\begin{figure}[!t]
  \centering
  \begin{tikzpicture}[
  participant/.style={
    draw, rectangle,
    minimum width=1.3cm, minimum height=0.45cm,
    font=\small\bfseries, inner sep=3pt
  },
  lifeline/.style={dashed, gray!70, thick, -{Stealth[length=5pt]}},
  msg/.style={-{Stealth[length=5pt]}, thick},
  lbl/.style={font=\footnotesize, inner sep=2pt},
  act/.style={font=\footnotesize, align=left, inner sep=0pt},
  col sep/.initial=3.2cm,
  row h/.initial=0.85cm,
]

  \def\ColSep{3.2cm}
  \def\RowH{0.85cm}
  \def\LLDepth{5.0cm}
  \node[participant] (B) {Broker};
  \node[participant, right=\ColSep of B] (S) {Source};

  \draw[lifeline] (B.south) -- ++(0,-\LLDepth);
  \draw[lifeline] (S.south) -- ++(0,-\LLDepth);

  \foreach \row in {1,...,5}{
    \coordinate (BL\row) at ($(B.south)+(0,-\row*\RowH)$);
    \coordinate (SL\row) at ($(S.south)+(0,-\row*\RowH)$);
  }

  \node[act, anchor=west] at ($(BL1)+(0.12,0)$)
    {\textcircled{\scriptsize 1}\ Pick unused challenge $(n_i,c_i)$};

  \draw[msg] (BL2) -- (SL2)
    node[midway, above, lbl] {\textcircled{\scriptsize 2}\ $(n_i,\; \mathcal{K}_{id})$};

  \node[act, anchor=east] at ($(SL3)+(-0.12,0)$)
    {\textcircled{\scriptsize 3}\ Compute $r_i=H(\mathcal{K}_{id}\|n_i\|W_{\Delta})$};

  \draw[msg] (SL4) -- (BL4)
    node[midway, above, lbl] {\textcircled{\scriptsize 4}\ $r_i$};

  \node[act, anchor=west] at ($(BL5)+(0.12,0)$)
    {\textcircled{\scriptsize 5}\ Verify:\ $r_i=c_i$};

\end{tikzpicture}
  \caption{A walk-through of the verification protocol. Figure converted to LateX using AI}
  \label{fig:verification-protocol}
\end{figure}

\noindent\textbf{Step 5: Query Execution.} The broker computes $R = \Delta(W_{\Delta})$.

\noindent\textbf{Step 6: Source Provenance Recording.} Before sending $R$ to the destination, the
broker must persist a mapping between the witness and the result $R$, $\prov(\Delta, W_{\Delta}, R$)
$\subseteq tid_{src}(I) \times tid(R)$, where $tid(R)$ are the temporary tuple ids of shared data.
In Figure~\ref{fig:sharing-protocol}, this is represented in the relation labeled \textbf{Prov}. At
this stage, the broker doesn't know the identifier $tid_{dst}(J')$, and must persist a temporary
mapping between tuple ids in the \textit{sharing ledger}, a broker-controlled relation that
documents data exchange events. There is a single ledger for each pair of sharing organizations,
where individual data exchange events are identified by $\mathcal{K}_{id}$.

\noindent\textbf{Step 7: Broker Sends Data to Destination.} The broker then sends the data $R$
(which includes $tid(R)$) to be shared to the destination, who ingests the data into their database.

\noindent\textbf{Step 8: Destination Sends Prov Mapping.} The destination must then return a mapping
between shared tuples and destination tuples to the broker. This mapping will be $\prov(\Delta, R,
J') \subseteq tid(R) \times tid_{dst}(J')$. On receipt, the broker updates the sharing ledger,
removing the temporary $tid(R)$, leaving the desired mapping $\prov(\Delta, I, J')$. The relation
\texttt{Prov} in Figure~\ref{fig:sharing-protocol} shows what the ledger looks like after a single
sharing event.

\noindent\textbf{Step 9: Broker Returns Receipt to Sender.} The final step of the sharing protocol
is for the broker to return a receipt of the data exchange event to the sender, which comprises a
hash of the shared data along with an ID for the event, $\mathcal{K}_{id}$. The \texttt{sender} can
use the hash to confirm that the data they intended to share was in fact the data that was shared.
The \texttt{sender} must also know $\mathcal{K}_{id}$ to identify the correct witness to use during
verification.

\subsection{Verification Protocol}\label{subsec:verification-protocol}
After sharing, the broker can verify up to $k$ times that the sharer has retained a witness using an
interactive proof. Figure~\ref{fig:verification-protocol} walks through each step of the
verification protocol.

Before walking through the verification protocol, we note that the sharer can choose to retain a
witness from among multiple candidate witnesses, all of which could satisfy the verification
protocol. Recall that a witness is a set of tuples sufficient to produce all the output tuples for a
given query. The sharer could satisfy every possible claim trivially by registering the entire
database as a witness. In this case, they promise to retain the database in its current form.
Usually, however, the \texttt{sender} will retain the subdatabase necessary to reproduce the
shared data.

One interesting choice of subdatabase is one that is a \textit{minimal witness},
or, a set of tuples that has no proper subset that is also a
witness~\cite{cheney2009provenance}. The minimal witness contains only tuples
that are necessary to produce the output. Some queries have multiple different
minimal witnesses, in particular certain SQL queries with \texttt{UNION} or
\texttt{AGG}. Although Datalog or recursive queries are out of scope for this
paper, a Datalog fact might be provable via multiple proof trees. To take a
simple example, a query asking whether two nodes A and B in a graph are
reachable (i.e., \texttt{reachable(A,B)}) could have multiple minimal witnesses
if there are multiple distinct paths from A to B. The fact that the sharer has
some flexibility to choose a witness means our framework is agnostic to the
physical data model used. Depending on the model in use, picking a minimal
witness may or may not be challenging. For example, row-based storage co-locates
all tuples attributes, while columnar-based storage each attribute is stored
separately. 

Figure~\ref{fig:verification-protocol}
illustrates the flow of
the verification verification protocol, which proceeds as follows. The broker
starts by picking an unused challenge ($n_i$, $c_i$). The broker then sends the
nonce $n_i$ and the contract id, $\mathcal{K}_{id}$, to the source. The source
then computes $r_i = H(\mathcal{K}_{id}\|n_i\|W_{\Delta})$, where $H$ is the
hash function used by the broker. The source return $r_i$ to the broker, which
confirms retention if $r_i= c_i$, and rejects otherwise. 

\subsection{Broker Implementation and API}\label{subsec:broker}
The broker is a trusted third party system whose implementation must meet the following
requirements to support the sharing and validation protocols. The broker API is given in
Figure~\ref{fig:broker-api}.

\afterpage{%
\begin{figure}[t]
\centering
\begin{tcolorbox}[
  enhanced,
  colback=white,
  colframe=black,
  boxrule=0.6pt,
  arc=0pt,
  left=5pt,
  right=5pt,
  top=4pt,
  bottom=4pt,
  before skip=6pt,
  after skip=8pt
]
\small
\textbf{Broker API}\par\smallskip
\noindent\texttt{Create}$(iam\_creds,\mathcal{O}_S,\mathcal{O}_R,\mathcal{V})
  \rightarrow \mathcal{K}_{id}$\par
Sender and destination register with their IAM credentials and establish a data exchange contract with
the broker.\par\smallskip

\noindent\texttt{Share}$(\mathcal{K}_{id},W_{\Delta},\Delta,P_{sat})
  \rightarrow (\mathcal{K}_{id},R)$\par
Broker validates $W_{\Delta}$ against $P_{sat}$, computes the challenges for verification, computes
$\Delta(W_{\Delta}) = R$, and sends $R$ to the destination.\par\smallskip

\noindent\texttt{Finalize}$(\mathcal{K}_{id},\prov(\Delta,R,J'))$\par
The destination sends a mapping between $R$ and its database, which the broker records as provenance.\par\smallskip

\noindent\texttt{Create\_Audit}$(\mathcal{K}_{id}) \rightarrow (\mathcal{K}_{id},c_i)$\par
Select an unused challenge nonce for contract $\mathcal{K}_{id}$, compute $c_i$, and open an audit.\par\smallskip

\noindent\texttt{Run\_Audit}$(\mathcal{K}_{id},c_i,a_i)
  \rightarrow \{accept,reject\}$\par
Verify audit $a_i$ for contract $\mathcal{K}_{id}$ using challenge $c_i$.
\end{tcolorbox}
\caption{Broker API. Figure converted to LateX using AI}
\label{fig:broker-api}
\end{figure}
}

To support this API, a minimal broker must be architected to support the
following. The necessity of sharing encryption keys for secure storage and
transmisison depends on the particular threat model. In the next section,
propose an implementation where the broker sits in an untrusted cloud service,
and therefore must encrypt all information before persisting or sending. 

\noindent\textbf{Witness and Challenge Computation.} To start, the broker must be able to take as
input a Parquet file for the witness (or a ZIP archive containing a set of parquet files), and a SQL query
file containing $\Delta$. Accordingly, the broker must be able to execute arbitrary SQL queries, and
have access to a limited shell for reading and writing files. To calculate and store challenges, the
broker must be able to compute hashes, and store and read those results from a database.

\noindent\textbf{Witness Validation}. The broker must also be able to validate that the provided
witness is valid and satisfies any additional constraints defined in $P_{sat}$. The ability to
execute SQL queries suffices to execute these two tasks.

\noindent\textbf{Secure Transmission}. Finally, the broker must be able to
transmit the result to the destination, receive the destination's mapping, and
send the receipt of the data exchange to the \texttt{sender}.

\section{Towards a Practical Implementation}\label{sec:threat-model}
In this section, we outline a system prototype and the trust assumptions we make about
each actor in the system. Then, we describe how the system could be set up on commodity cloud
infrastructure.

\subsection{Practical Threat Model}

\textbf{Actors}. There are four actors in a given sharing event: the \textit{source} that shares
data with a \textit{destination} via a trusted \textit{broker} $B$. To share data, the source sends the
broker a query $\Delta$, witness $W_{\Delta}$, and (optionally) a set of claims $P_{sat}$. The
broker then verifies the claims, executes $\Delta(W_{\Delta})$, computes and stores the audit
challenges, and finally sends the result $\Delta(W_{\Delta})$ to the destination $R$. Although only one logical broker is needed for each exchange, a single broker service can mediate many
such pairs. The broker may sit in independently run commodity cloud infrastructure owned by a 
an actor we refer to as the \textit{cloud operator}.

\textbf{Trust assumptions}. The source and destination actors may be mutually distrusting. We
discuss how each of them could disrupt the sharing or verification protocol in this section, and how
our system mitigates some of these threats. We do, however, assume that both the source and
destination trust the broker to securely and correctly mediate the sharing event. Securely means
that the broker doesn't egress or persist inputs ($W_{\Delta}$, $\Delta$, and $P_{sat}$),
intermediate results, or outputs ($\Delta(W_{\Delta})$). The broker only ever persists the audit
challenges, the provenance ledger, and other sharing metadata. Correctly means that the broker
correctly executes the procedure (i.e. as opposed to running a different function).

\textbf{Security Goals}. Our system is designed to achieve two goals. First, an actor that has not
retained a witness should not pass an audit, except with negligible probability. Second, the
unencrypted inputs of the exchange should only ever be persisted by the source and the
unencrypted outputs only persisted by the destination.

\subsubsection{Threats from the Source}
A malicious source could attempt to disrupt the sharing or verification protocol in the
following ways:

\textbf{Sharing a fake witness}. Only the source will have access to their database,
$\mathcal{D}_{src}$. Thus, a malicious source could in principle register a witness $W_{\Delta}$
that does not accurately reflect $\mathcal{D}_{src}$. In particular, the source agent might share
$W_{\Delta}$ such that either ($\Delta(W_{\Delta}) \neq \Delta(\mathcal{D}_{src})$) or ($W_{\Delta}
\nsubseteq \mathcal{D}_{src}$). Unless the broker has access to $\mathcal{D}_{src}$, it's not
possible to identify a fake witness. We must therefore assume that the source sends a witness that
satisfies both.

Based on our problem statement, both of these assumptions are reasonable. A source that shares a
witness that is not responsive to the query $\Delta$ (e.g. by omitting rows or including ``fake''
data) is in violation of the data exchange contract. But doing so will not give rise to a
provenance anomaly because our framework will accurately record the exchange that happened, even if
the data shared is not what the contract specified. Likewise, sharing a witness that is not a
subdatabase of $\mathcal{D}_{src}$ would subvert destination expectations, but would not produce a
provenance anomaly. In either case, the sharer could be punished by either the destination or a third
party, in the form of exclusion from future sharing, or regulatory penalty. But such non-technical
incentives are outside the scope of this work.

\textbf{Discarding a registered witness}. The source could thwart the retention requirements by
discarding the registered witness. In this case, however, an audit will reveal that the source has
failed to retain the witness because, with high probability, they will not be able to respond to the
challenge in the verification protocol. Although there is no mechanism to prevent deletion, having
the ability to verify retention is sufficient to satisfy our goal of detecting when one of the
parties do not satisfy one of the contract's provisions.

\textbf{Forge audit checks}. The likelihood that the source could successfully respond to a
verification challenge without retaining the witness is a function of the broker's hash function.
Recall the audit mechanism: the broker precomputes $h_i = \texttt{hash}(W_{\Delta} || n_i )$ for
some nonce $n_i$ at sharing time, then challenges the source to compute $h_i$ given $n_i$.
\texttt{hash} is a collision-resistant hash function from $\{0, 1\}^l \to \{0, 1\}^{\lambda}$, where
$l$ is the variable length input and $\lambda$ is a fixed-length hyperparameter. In this setup, the
probability that the source chooses the correct $h_i$ without access to $W_{\Delta}$ is
$(\frac{1}{2})^{\lambda}$. The broker can thus pick an appropriate $\lambda$ that makes the
likelihood of guessing $h_i$ vanishingly small.

\subsubsection{Threats from the Destination}
Because the destination plays a mostly passive role in data sharing, there is not much an
adversarial agent can accomplish.

\textbf{Record mapping forgery}. When the destination ingests the result $R=\Delta(W_{\Delta})$, the
sharing protocol requires them to send, for each tuple $t\in R$, a mapping between the primary key
of $\Delta(W_{\Delta})$ and the primary key of the destination relations. Executing forward and backward
provenance queries across multiple data exchanges is only possible if this mapping is accurate. A
malicious destination could send a false mapping, but the broker could detect this through the same
verification protocol it uses to check that the source has retained the witness, only this time
using $\Delta(W_{\Delta})$ instead of $W_{\Delta}$ in the challenge. Our current protocol does not
currently require the broker to compute challenges for the destination but it could easily be
extended to do so.

\subsubsection{Threats from the Broker}
Our design makes the following assumptions about the trustworthiness of the broker, including:
\begin{enumerate}[leftmargin=*, label=\arabic*.]
  \item \textit{Broker correctly verifies the witness}. One threat is that the broker colludes with
    the source, allowing them to register a witness that is either not sound or not valid.
    To address this threat, the framework is designed to support the use of open-source and
    trusted broker software that each party can verify (more on this in Section~\ref{subsec:reference-implementation}).
    So long as this is the case, we can assume that the broker and source cannot collude.
  \item \textit{Broker doesn't egress sensitive information}. At sharing time, the broker is trusted
    to possess $\Delta$, $W_{\Delta}$, and $P_{sat}$, as well as to compute $\Delta(W_{\Delta})$. We
    assume that the broker only sends $\Delta(W_{\Delta})$ to the destination. To minimize the risk
    that sensitive information leaks, the broker only ever stores and computes sensitive information
    in-memory, and persists only the audit hashes, nonces, primary-key mappings, and other metadata.
    The next section discusses how to implement the framework in this way.
  \item \textit{Broker persists information}. The broker only persists metadata about the shared
    data, including the query, claims, and other information to link the source and destination.
\end{enumerate}

The primary threat that we are not robust to is \emph{broker compromise}. If a third party accesses
and controls the broker through an existing security vulnerability, our framework cannot ensure that
provenance anomalies will not occur. Thus, we must assume that the broker is not compromised.

\subsection{AWS Reference Implementation}\label{subsec:reference-implementation}
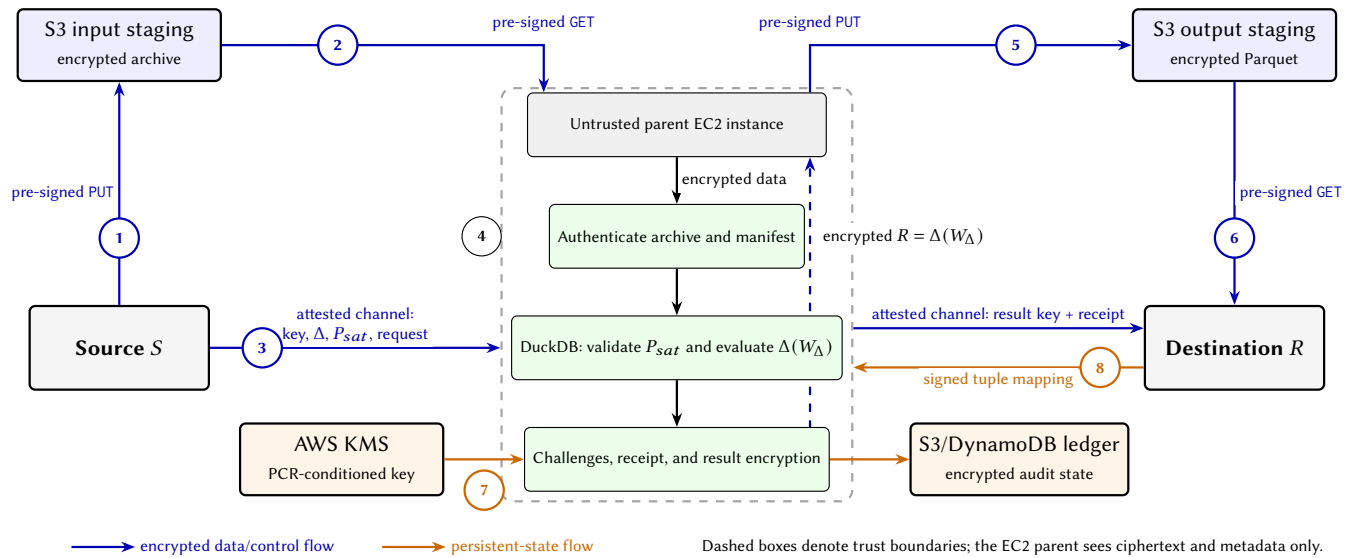
\begin{figure*}[t]
  \centering
  \resizebox{\textwidth}{!}{\begin{tikzpicture}[
  font=\sffamily,
  >=Stealth,
  actor/.style={
    draw, rounded corners=2pt, thick, fill=gray!8,
    minimum width=2.25cm, minimum height=1.05cm,
    align=center, font=\small\sffamily\bfseries
  },
  service/.style={
    draw, rounded corners=2pt, thick, fill=blue!7,
    minimum width=2.55cm, minimum height=0.9cm,
    align=center, font=\small\sffamily
  },
  component/.style={
    draw, rounded corners=2pt, fill=green!8,
    minimum width=3.0cm, minimum height=0.8cm,
    align=center, font=\scriptsize\sffamily
  },
  flow/.style={-{Stealth[length=5pt]}, thick},
  cryptoflow/.style={flow, blue!70!black},
  stateflow/.style={flow, orange!80!black},
  boundary/.style={draw, dashed, thick, rounded corners=4pt, inner sep=8pt},
  stepnum/.style={
    circle, draw, fill=white, inner sep=1pt, minimum size=14pt,
    font=\scriptsize\sffamily\bfseries
  },
  label/.style={font=\scriptsize\sffamily, align=center, fill=white, inner sep=1.5pt}
]

% Participants and cloud services.
\node[actor] (source) at (-7.0,-0.6) {Source $S$};
\node[actor] (destination) at (7.0,-0.6) {Destination $R$};
\node[service] (s3in) at (-7.0,3.2) {S3 input staging\\{\scriptsize encrypted archive}};
\node[service] (s3out) at (7.0,3.2) {S3 output staging\\{\scriptsize encrypted Parquet}};
% Parent instance and enclave.
\node[component, fill=gray!12, minimum width=3.75cm] (gateway) at (0,2.2)
  {Untrusted parent EC2 instance};
\node[component] (verify) at (0,0.8)
  {Authenticate archive and manifest};
\node[component] (duckdb) at (0,-0.6)
  {DuckDB: validate $P_{sat}$ and evaluate $\Delta(W_{\Delta})$};
\node[component] (commit) at (0,-2.0)
  {Challenges, receipt, and result encryption};
\node[service, fill=orange!9, anchor=east] (kms)
  at ([xshift=-1.0cm]commit.west)
  {AWS KMS\\{\scriptsize PCR-conditioned key}};
\node[service, fill=orange!9, anchor=west] (ledger)
  at ([xshift=1.0cm]commit.east)
  {S3/DynamoDB ledger\\{\scriptsize encrypted audit state}};

\begin{scope}[on background layer]
  \node[boundary, fill=green!3, fit=(verify)(duckdb)(commit),
        label={[font=\scriptsize\sffamily\bfseries]above:Nitro Enclave (attested open-source EIF)}]
        (enclave) {};
  \node[boundary, draw=gray!70, fit=(gateway)(enclave),
        label={[font=\scriptsize\sffamily\bfseries]above:EC2 parent instance}]
        (parent) {};
\end{scope}

% Numbered data flows.
\draw[cryptoflow] (source.north) -- (s3in.south)
  node[midway, left, label] {pre-signed \texttt{PUT}}
  node[pos=0.30, stepnum] {1};

\draw[cryptoflow] (s3in.east) -| ([xshift=6pt]gateway.north west)
  node[pos=0.55, above, label, yshift=5pt] {pre-signed \texttt{GET}}
  node[pos=0.18, stepnum] {2};

\draw[cryptoflow] (source.east) -- (parent.west |- source.east)
  node[midway, above, label] {attested channel:\\key, $\Delta$, $P_{sat}$, request}
  node[pos=0.18, stepnum] {3};

\draw[flow] (gateway.south) -- (verify.north)
  node[midway, right, label] {encrypted data};

\draw[flow] (verify.south) -- (duckdb.north);
\draw[flow] (duckdb.south) -- (commit.north);
\node[stepnum] at ([xshift=-24pt]verify.west) {4};

\begin{scope}[on background layer]
  \draw[cryptoflow, dashed]
    (commit.east) -- ([xshift=-6pt]gateway.east |- commit.east)
    -- ([xshift=-6pt]gateway.south east);
\end{scope}
\node[label, anchor=west] at ([xshift=-3pt]gateway.east |- verify.east)
  {encrypted $R=\Delta(W_{\Delta})$};

\draw[cryptoflow] ([xshift=-6pt]gateway.north east) |- (s3out.west)
  node[pos=0.45, above, label, yshift=5pt] {pre-signed \texttt{PUT}}
  node[pos=0.82, stepnum] {5};

\draw[cryptoflow] (s3out.south) -- (destination.north)
  node[midway, right, label] {pre-signed \texttt{GET}}
  node[pos=0.70, stepnum] {6};

\draw[cryptoflow] ($(parent.east |- duckdb.east)+(0,7pt)$)
  -- ([yshift=7pt]destination.west)
  node[midway, above, label] {attested channel: result key + receipt};

\draw[stateflow] (kms.east) -- (commit.west)
  node[midway, stepnum, yshift=-11pt] {7};
\draw[stateflow] (commit.east) -- (ledger.west);

\draw[stateflow] ([yshift=-7pt]destination.west)
  -- ($(parent.east |- duckdb.east)+(0,-7pt)$)
  node[midway, below, label] {signed tuple mapping}
  node[pos=0.15, stepnum] {8};

% Legend.
\draw[cryptoflow] (-7.6,-3.1) -- ++(0.8,0)
  node[right, label, fill=none] {encrypted data/control flow};
\draw[stateflow] (-3.7,-3.1) -- ++(0.8,0)
  node[right, label, fill=none] {persistent-state flow};
\node[font=\scriptsize\sffamily, anchor=west] at (0.2,-3.1)
  {Dashed boxes denote trust boundaries; the EC2 parent sees ciphertext and metadata only.};

\end{tikzpicture}}
  \caption{Reference implementation and data flow. The source uploads an encrypted archive of
  Parquet files using a pre-signed S3 URL (1). The untrusted EC2 gateway retrieves and relays the
  encrypted data (2), while secrets and the signed request enter over a channel bound to the enclave's
  attested key (3). The enclave authenticates the archive and evaluates $P_{sat}$ and query in
  DuckDB (4), then encrypts and uploads the result $R=\Delta(W_{\Delta})$ (5). The destination downloads and verifies $R$ (6).
  Provenance is encrypted using an attestation-conditioned KMS key (7), and the destination returns its
  signed tuple mapping (8). Figure converted to LateX using AI}
  \label{fig:practical-implementation}
\end{figure*}

We next provide a concrete reference architecture for the
broker implemented with standard tools in AWS, illustrated by
Figure~\ref{fig:practical-implementation}.
This reference architecture is not required by the protocol. S3 can be replaced
by another object store, DuckDB by another relational engine, and Nitro Enclaves
by any TEE that provides isolation and remote attestation. This architecture is
to make explicit the data formats, trust boundaries, and failure conditions
needed by an implementation.

\subsubsection{Deployment and data format}
The broker service runs on a trusted Nitro Enclave that sits inside an untrusted parent EC2
instance. The parent instance is responsible for authenticating parties, creating jobs for the
broker, and relaying pre-signed information between the parties and the broker. The parent instance
does not process queries or hold cryptographic keys. The enclave, on the other hand, must contain
everything necessary to execute the protocols, including a protocol state machine, a SQL parser and
policy checker, DuckDB, the ability to read Parquet files, and the code used to construct challenges
and update the ledger. The enclave can only perform computation. Any external communication or
access to persistent storage is mediated by the parent.

The witness $W_{\Delta}$ is an archive containing a manifest and one or more Parquet files. The
manifest includes identifiers for both parties and the exchange event, as well as logical metadata
for $W_{\Delta}$ (relation names, schemas, primary- and foreign-key declarations, row counts) and
integrity checks for the archive (an ordered list of archive members, a cryptographic digest, and
byte length for every member). For a given data exchange contract, the source submits a
signed job request that includes the archive, and binds the manifest to the other elements of the
contract: $\Delta$, $P_{sat}$, the destination, and the unique contract identifier,
$\mathcal{K}_{id}$. To avoid ambiguous or duplicative encodings, the archive must not have (e.g.)
absolute paths, links, or undeclared members. To compute challenges, the broker hashes the elements of
the exchange in a predetermined order, as opposed to the database's potentially unstable row order.

\subsubsection{Encrypted transmission and attested execution}
To start, the source must make the archive accessible to the broker. The broker sends a
short-lived, pre-signed PUT URL that points to such a publicly inaccessible S3 staging bucket to the
source. The source can then upload the encrypted archive to the S3 staging bucket,
once again, using a predetermined encryption scheme. Although the URL will be revealed to the
parent, the stored archive is encrypted, so they can only learn the upload time and size of archive.
Once uploaded, the source submits a signed job request to the broker, and the broker
verifies the archive against the manifest. S3 versioning or a write-once random key prevents a later
upload from silently replacing the accepted input.

Next, the source and destination must verify the trustworthiness of the enclave. To
do so, the broker generates a temporary public key, and asks the Nitro Enclave for an attestation
document that contains that key, as well as a party-provided nonce to prevent the Enclave from
sending an old attestation. The source and destination first verify the attestation,
then use the broker-provided key to encrypt and send the signed job request, $\Delta$, $P_{sat}$,
and the key that decrypts the archive. On receipt, the broker fetches the encrypted archive using a
pre-signed GET URL. The broker then decrypts the archive and performs integrity checks before
evaluating a query. In this scheme, the parent instance is merely a byte relay. Although the parent
may drop packets or issue a denial of service, it cannot read or undetectably modify $W_{\Delta}$.

Once the archive is decrypted and authenticated, the broker unpacks it into the enclave memory. To
compute the result and challenges, the broker only needs access to a minimal, in-memory query engine
that disallows actions that seek to compromise the broker's responsibilities, such as running
nondeterministic functions, accessing another external network, installing extensions, or attempts
to change the broker's configuration. In this reference implementation, the broker uses a DuckDB
instance modified to meet these requirements. In particular, the broker must validate that
$W_{\Delta}$ satisfies $P_{sat}$ (rejecting the query on failure), compute $R=\Delta(W_{\Delta})$, and
construct the challenges 
\begin{align*}
  c_i=H(\mathcal{K}_{id}\|n_i\|W_{\Delta}).
\end{align*}

for $i$ in $k$ challenges, as specified by the contract. Including the contract identifier in
challenge creation prevents a response from one exchange from being reused for another. To ensure
consistency across time, the broker pins the Parquet and DuckDB versions used. Otherwise,
software updates may render verification impossible.  

The broker's DuckDB instance must be further modified to protect against threats that can occur
during computation. First, the broker must place per-contract memory and CPU use limits on query
computation to protect against denial of service attacks. Second, for certain query operations that
require more memory than is currently available (e.g. joins or nested queries), databases will
temporarily persist intermediate results to disk. But persisting unencrypted results could
potentially leak information to the parent instance. In this implementation, we disable such
temporary writes, although an alternative solution would require the broker to encrypt intermediate
results with a temporary key and erase any data once computation has completed. 

If every check and computation succeeds, the enclave produces the encrypted archive containing
Parquet files for $R$ and the result manifest, as well as a signed receipt binding the input and
result digests, $\Delta$, $P_{sat}$. The broker doesn't publish the result on partial failure. At
this point, the broker will write the first half of the exchange to the provenance ledger, which is
persisted in S3/DynamoDB. To make these results accessible to the destination, the broker creates a
temporary encryption key, obtains from the parent a pre-signed PUT URL pointing to a temporary S3
bucket, and uploads the encrypted results to that bucket. To communicate with the destination, the
broker establishes an attested channel with the destination, following a similar procedure as it did
to communicate with the source. Once established, the destination receives a short-lived pre-signed
\texttt{GET} URL for the S3 bucket, as well as the decryption key. On receipt, the destination
verifies the key, digest, and signed receipt and ingests the Parquest files containing the result.

Finally, the destination returns to the broker a signed mapping from the shared tuple IDs and its
local tuple IDs. The broker records that mapping in the provenance ledger and acknowledges
completion. After a timeout or source and destination acknowledgment, the broker then deletes and
objects left in the S3 buckets used for staging. This deletion serves an operational purpose, and is
not a necessary to preserve privacy because staged objects are encrypted.  

\subsubsection{Persistent state and reproducible trust}

Only challenges and their use status, nonces, provenance mappings, protocol metadata, digests, and
signed receipts survive the exchange job. We take two precautions to protect this information. To
start, the provenance ledger is an append-only hash chain. That way, the broker can detect rollback
or deletion by the cloud operator. Second, all of this information is encrypted using a specific key per
exchange, where keys are managed using the AWS Key Management System (KMS). The AWS KMS permits a
policy that allows the broker to access data encryption keys upon presenting a valid attestation
document that contains an approved broker measurement (e.g., PCR0, and PCR8 when signed enclave
images are used). If approved, the KMS uses the broker-supplied public key to encrypt the key to
encrypt the information to be persisted, and sends the resulting ciphertext to the broker via the parent.  
The encrypted data may then be stored in S3 or DynamoDB. 

The parties need not trust the broker service because the broker image is to be made publicly
available, including the container definition, locked dependency versions, build scripts, and
anything else necessary to repodicibly rebuild the image. That way, parties can verify the
open-source broker before use, as well as compare its measurement with the value in the attestation
document and KMS policy.  Although this construction narrows the trust that the parties must place
in the broker, it doesn't eliminate trust. In particular, the parties must trust the enclave runtime
and AWS's Nitro and KMS implementations. There also remains the possibility of hitherto unknown but
exploitable side channels. Traffic analysis, denial of service, a compromised source or destination,
and physical attacks remain outside the guarantees of this reference architecture.

\section{Related Work}
In this section, we contextualize our work within related work studying provenance and proof of
possession and correctness to further motivate \textit{Proof-of-Retention}.

\subsection{Provenance}

A number of works have sought to trace and secure the provenance (or lineage) of data as it is
shared~\cite{backes2014lime, zhang2019auditshare, kho2015design} or undergoes
transformations~\cite{ruan2021lineagechain, park2011ramp, hasan2009preventing, liang2017provchain,
mast2018enabling}. None of these works, with the exception of the two we will discuss, are designed
with the goal of tracking and enforcing provenance as data is shared across control boundaries.
Within a single organization, maintaining provenance is simpler because all data transformation,
migrations, and updates are centrally controlled. Across control boundaries, however, sharing
parties can make unilateral changes to their data, potentially rendering the provenance stale. Thus,
a cross-organization system must provide a coordination or retention mechanism to prevent the
provenance record from becoming stale.

The most relevant related works for our purposes are the \texttt{ORCHESTRA} data
sharing system~\cite{ives2008orchestra} and the Data Station~\cite{xia2023data}. \texttt{ORCHESTRA}
collaborative data sharing system is a framework for exchanging data between autonomous peers. Each
participant controls a local database, and can optionally publish and fetch updates to peers in the
same sharing network. Updates are applied using schema mappings between each participant.s
databases, enabling each ingested tuple's provenance to be captured. But unlike our framework,
\texttt{ORCHESTRA} does not provide a system to ensure that the captured provenance is not stale, which
can occur if the origin updates their database but does not publish to peers.
In addition, \texttt{ORCHESTRA} does not support a mechanism to enforce retention of data that is
related to, but ultimately not shared. For example, firms sharing data subject to the CPRA can be
required to retain users' opt-out choice under Proof-of-Retention without sharing that information.

The Data Station~\cite{xia2023data} similarly focuses on facilitating data sharing, but does not
define retention obligations nor enforce those obligations. In particular, Data Station operations
are logged in terms of pointers or identifiers to data. But like \texttt{ORCHESTRA}, the log can go
stale if the sharing party unilaterally changes the data pointed to by the log.
\textit{Proof-of-Retention} requires (and enforces) senders to maintain a witness for shared data.

Tracking and analyzing data provenance is also a common approach auditing
databases and data sharing. Pasquier et al. study the challenges of auditing
data flows between IoT devices and propose using, among other representations, a
data provenance DAG~\cite{pasquier2018data}. Bier discusses how well provenance
tracking can fulfill and support data protection
goals~\cite{bier2013usage}. Other works have studied the appropriate
representation of provenance to support auditing~\cite{butin2016formal}. 

\subsection{Proof of Possession and Correctness}

The main shortcoming of the provenance-related works just discussed is that they
lack a mechanism to ensure that the provenance log does not become stale as a
result of unilateral action. This desired mechanism is closely related to two
lines of work: proof of possession/retrievability and verifiable database
operations.

In proof of possession works, the goal is to design a verification scheme for a
client to check whether data stored in an untrusted location is still
present~\cite{ateniese2007provable, shah2008privacy, halevi2011proofs}. In general, these schemes
require the client to pre-compute a challenge based on the stored data, and ask
the untrusted location to answer or reproduce the challenge at some future
point. Extensions to the basic scheme include updates to stored
data~\cite{erway2015dynamic, shi2013practical} and proofs of full
retrievability~\cite{juels2007pors,shacham2013compact}.

Similarly, related work studying proofs of correctness for queries over
untrusted databases is relevant to \textit{Proof-of-Retention}. In general, the
goal of query correctness proofs is to demonstrate to the challenger that the
query result is correct, without giving the challenger access to the (dynamic)
source database (and sometimes participant identity~\cite{wang2012knox,
wang2014oruta}), with the exception of a setup stage~\cite{zhang2015integridb,
zhang2017vsql,gu2025poneglyphdb, li2023zksql}. Although
\textit{Proof-of-Retention} uses a similar interactive proof structure, none of
the proof-of-possession, retrievability, or correctness schemes connect retained
data to specific cross-organization data exchange events. Moreover, the
proof-of-possession literature focuses on verifying outsourced data storage.

\section{Discussion}
In this section, we discuss how \textit{Proof-of-Retention} relates to other privacy concepts, how
and where it can be best used, and how witness-based retention requirements can improve existing
legal retention requirements.

\subsection{Relation to Data Minimization}
At first blush, our proposal to require entities to retain more data than they
otherwise would seems at odds with the principle of data minimization. Indeed,
other frameworks start from the assumption that databases will be subject to
data retention \textit{limitations}, that is, affirmative restrictions on what
data entities are permitted to preserve~\cite{lu2013auditing}, and develop
mechanisms to audit database in the absence of data. 

Our work, however, addresses a different privacy harm than that targeted by data
minimization~\cite{sharma2024m}. For context, the goal of data minimization is
to limit the collection or acquisition of data to only that which is strictly
necessary (see e.g. Article 5(1)(c) of the GDPR~\cite{gdpr2016art5}). Of course,
the principle of data minimization can be applied to data sharing, not just data
collection. But our goal is to ensure that any given data exchange, once it has
already occurred, is still auditable. To vindicate this goal, our framework
ensures that the information necessary to determine whether a particular
exchange was appropriate or privacy-respecting is preserved. In other words,
\textit{Proof-of-Retention} is a mechanism to determine how well entities
achieve data minimization during data sharing.  

\subsection{Generalizing Data Retention}
The purpose of a data retention obligation is to make future queries answerable. For example,
when FERPA requires an educational institution to maintain a record of every disclosure of
personally identifiable information, the goal is to be able to answer future questions about (e.g.) when
and what data was disclosed, and to whom.

Most data retention obligations today are defined in terms of the data to be
preserved, rather than the questions to be answered. We can refer to this common
type of data retention requirement a \textit{data-centric} obligation. This
approach makes sense when the future questions are not known ex ante. Consider
SEC Rule 17a-4, which requires broker-dealers to retain all ``business
communication and records.'' The purpose of this rule is to ensure that
investigators or auditors have access to the information necessary to
investigate and identify wrongdoings. Most of the time, however, it's not
possible to know which records will be necessary or relevant to an investigation
ahead of time because we don't know who or when crimes will be committed. Thus,
Rule 17a-4 must be data-centric. 

But there exists an alternative to a data-centric requirement:
\textit{query-centric} obligations. Rather than defining obligations in terms of
the data's content, we define them in terms of the future questions they must
answer. For example, rather than require financial firms to retain records of
transactions and customers, a query-centric retention obligation would take the
form ``retain all records necessary to determine which transactions occurred and
the parties involved.''

For some retention requirements, the data-centric and query-centric versions
will appear identical in substance. But the query-centric framework actually
shifts the burden of deciding what data should be saved from the regulator to
the data owner, and can lead to more precise retention. To see why, consider
what the rulemaking body does when they design a data-centric obligation. They
start with a set of questions that they want to be answerable and select the
data they think will be useful for answering those questions. But, as is
generally assumed, private firms, not the government, have the best
understanding of their operations and data. Thus, the government's choice of data is more likely
than the firm's choice to be underinclusive (i.e. not able to answer the
questions), or overinclusive (i.e. more data than necessary to answer the
questions). Designing a query-centric obligation helps eliminate this
imprecision.

The following two use cases illustrate how a query-centric obligation can be
preferable to a data-centric one.

\textbf{Avoiding redundancy}. When a regulation establishes content-based retention requirements, the
data owner may end up retaining multiple copies of the same data. Consider again the SEC rule
requiring broker-dealers to preserve all ``business communication'' records. If there is overlap in
the content of these records, a content-based retention requirement would lead to redundant data
retention, and a consequent failure to achieve data minimization.

\textbf{Government Records and FOIA}. The US Federal Records Act establishes the ``legal framework for
federal records management, including record creation, maintenance, and disposition.'' The Act
authorizes agencies to develop record retention schedules. Typically, these schedules define
retention obligations based on the content of the records. But content-based retention obligations
can become outdated. Suppose a government agency makes a decision (e.g. disbursement of aid) using an
algorithmic or automatic system and we would like to preserve the records or data used to make that
decision. Content-based requirements (e.g. training data, input, weights, etc) would have to be updated
any time the decision-making process changes. Defining the obligation in terms of the question
ensures that there is no gap between the obligation and the question to be answered.

\textit{Proof-of-Retention} is a first step towards defining query-centric retention obligations.
Although this work focuses on ensuring retention for the purpose of auditing data exchanges,
requiring parties to preserve a witness is the basis of query-centric retention. Shifting the
retention requirement from specified data elements to a witness more closely aligns the data that is
preserved with the data that should be preserved.

\section{Acknowledgements}
Section omitted during review. 

\section{Ethical Considerations}
This work does not raise any ethical considerations. 

\section{Open Science}
This work did not produce any research artifacts. 

\section{AI Use}
For each figure in this paper, we first sketched out the layout and content by
hand, and then used ChatGPT (Codex) to convert that sketch into a LateX figure.
We also used Codex to format latex macros. We also used Codex to check grammar. 

We use AI Deep Research functionality to search for related work and
documentation for AWS infrastructure.

For all AI uses, we have manually verified and are responsible for the accuracy,
originality, and integrity of those uses.

\bibliographystyle{ACM-Reference-Format}
\bibliography{sample-base}

\end{document}